\documentclass[epj]{webofc}
\woctitle{MESON2018 - the 15$^\textrm{th}$ International Workshop on Meson Physics}
\begin{document}
\selectlanguage{english}
\title{A study of the vector meson $\mathbf{\psi(4040)}$}

\author{M. Piotrowska\inst{1}\fnsep\thanks{\email{milena.soltysiak@op.pl}} \and
        F. Giacosa\inst{1,2}
}

\institute{Institute of Physics, Jan Kochanowski University,\\\textit{ul. Swietokrzyska 15, 25-406, Kielce, Poland. }\\ 
\and
           Institute for Theoretical Physics, J. W. Goethe University,\\\textit{ Max-von-Laue-Str. 1, 60438 Frankfurt, Germany.}
          }

\abstract{We investigate the well-known $c\bar{c}$ vector state $\psi(4040)$ in the framework of a quantum field theoretical model. In particular, we study its spectral function and search for the pole(s) in the complex plane. Quite interestingly, the spectral function has a non-standard shape and two poles are present. The role of the meson-meson quantum loops (in particular $DD^*$ ones) is crucial and could also explain the not yet confirmed ``state'' $Y(4008)$.
}
\maketitle
\section{Introduction}

In the past 15 years many new mesonic states have been observed in the charmonium sector. However, not all of them can be accommodated in the picture of the standard quark model, where mesons are conventional quark-antiquark ($q\bar{q}$) objects. Some of the newly discovered states, called X, Y and Z, are good candidates for non-conventional mesons such as hybrids, multiquarks, molecules or glueballs, see Refs. \citep{rev, rev2016} and refs. therein.

In these proceedings, based on the forthcoming paper \cite{newpaper}, we explore the energy regime close to $4$ GeV, and we concentrate on the vector state $\psi(4040)$ \cite{pdg2018}. This resonance is a conventional $c\bar{c}$ state characterised by the quantum numbers  $n$ $^{2S+1}L_{J}=$ $3$~ $^{3}S_{1}$, where $n, S, L, J$ are the principal number, the spin, the angular momentum and the total spin, respectively. Moreover, in the charmonium spectrum, at about 4 GeV (very close to $\psi(4040)$) a puzzling broad enhancement, called $Y(4008)$, was noticed by the Belle Collaboration when studying the cross-section of $e^+ e^- \rightarrow J/ \Psi \pi^+ \pi^-$ \cite{belle, belle2}. However, this ``state'' was not confirmed by other groups measuring the same process. Although not yet confirmed, there have been various theoretical speculations (mostly within non-conventional scenarios) trying to explain its nature \cite{rev2016}. In our approach, we checked if it is possible to describe both $\psi(4040)$ and $Y(4008)$ simultaneously within a model in which only a single ordinary $c\bar{c}$ seed state corresponding to $\psi(4040)$ is taken into account. A similar method was used to study the scalar kaon $K^*_0(1430)$, where an additional companion pole describing the light $\kappa$ emerges \cite{milenathomas}. 
\section{Theoretical model and results}
First, we introduce the following relativistic interaction Lagrangian:
\begin{equation}
\mathcal{L}=ig_{\psi DD}\psi_{\mu}\left(\partial^{\mu}D^+D^-\right)+ig_{\psi D^*D}\tilde{\psi}_{\mu \nu}\left(\partial^{\mu}D^{*+ \nu}D^-\right)+ig_{\psi D^* D^*}\psi_{\mu \nu}\left(D^{*+ \mu}D^{*- \nu}\right)+h.c.\text{ }. \label{Lag}
\end{equation}
Each term describes different decays of the state $\psi(4040)$: $DD$ (and also $D_s D_s$) in the first term, $D^* D$ (and $D^*_s D_s$) in the second term, and $D^* D^*$ in the last one. In our model we have five free parameters: the bare mass $M_{\psi}$ of the state $\psi$, the three coupling constants $g_{\psi DD}$, $g_{\psi D^*D}$ and $g_{\psi D^* D^*}$ in Eq. (\ref{Lag}) (determined by using some results listed in the PDG \cite{pdg2018}), and finally, the cutoff $\Lambda$. The latter is part of the cutoff function (or form factor) $F_{\Lambda}$, which assures that our model is finite. We test two types of the form factor, the Gaussian and the dipolar one, see Table 1. Here, for illustration we use $\Lambda=0.42$ GeV, for which the pole width is exactly 80 MeV in the Gaussian case. The numerical values of the parameters are presented in the second column of Table 1. 

 Next, we evaluate the spectral function (alias the mass distribution) of the resonance $\psi(4040)$. In Fig. 1.a we show the results for both types of the cutoff function. It is visible that, besides the expected peak close to $4.04$  GeV, an additional enhancement appears close to $3.9$ GeV. In both cases, the shape of the spectral function has not a standard Breit-Wigner form, mainly due to the presence of a deformation left from the peak generated by mesonic quantum loops (most notably $DD^*$ ones).

\begin{figure}[h]
\begin{center}
\begin{minipage}[b]{6.90cm}
\centering
\includegraphics[width=6.50cm]{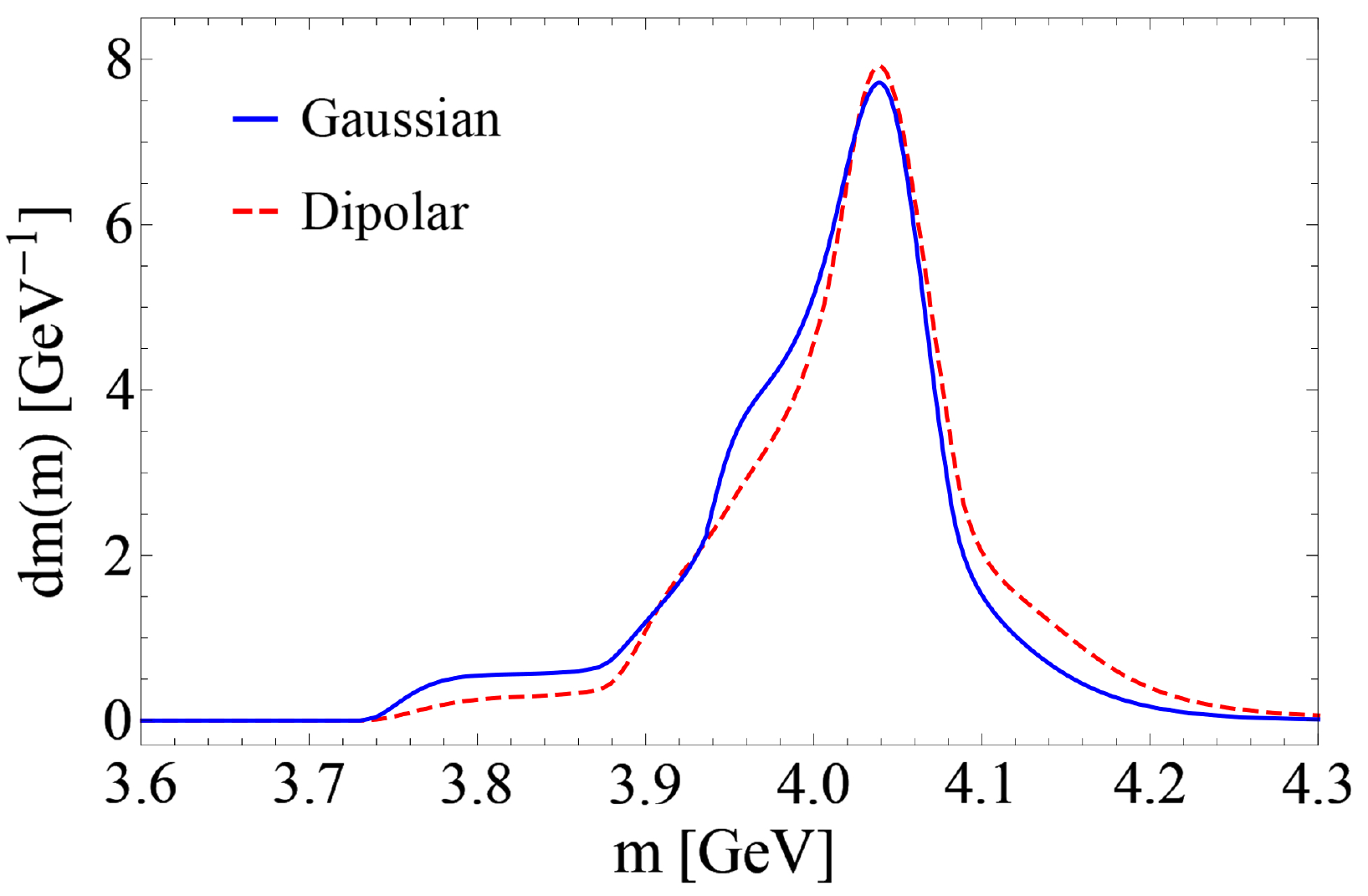}\\\textit{a)}
\end{minipage}
\begin{minipage}[b]{6.50cm}
\centering
\includegraphics[width=6.80cm]{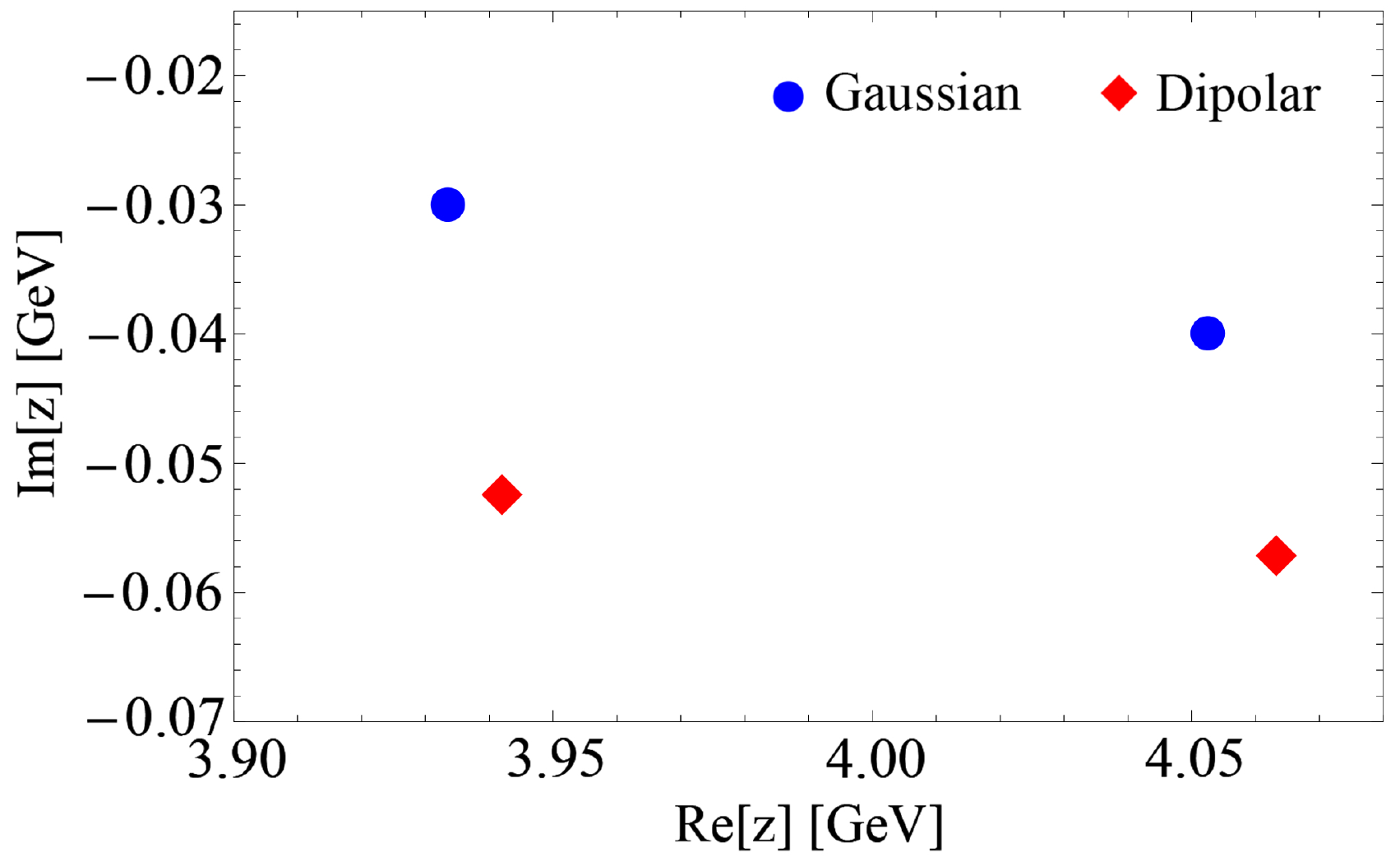}\\\textit{b)}
\end{minipage}
\end{center}
\caption{Panel (a) shows the form of the spectral function of the state $\psi(4040)$ for a Gaussian form factor (blue solid line) and for dipolar one (red dashed line). Panel (b) shows the position of the poles in the complex plane. Blue dots corresponds to the Gaussian form factor, while the red marks to a dipolar one.}%
\label{rys1}%
\end{figure}

 Moreover, even if a single seed state is present, we find that two poles appear in the complex plane, one related to the state $\psi(4040)$ and the second to the enhancement. The results for these poles are shown in Fig. 1.b and in the last two columns of Table 1.
\begin{table}[h!] 
\centering
\renewcommand{\arraystretch}{1.25}
 
\begin{tabular}[c]{cccc}
\hline
Form factor & Parameters & Pole for $\psi(4040)$& Second pole\\
\hline
$F_{\Lambda}=e^{-2k^2/ \Lambda^2}$&$g_{\psi DD}=39.6 \pm 5.0$&$(4.053 \pm 0.004)$&$(3.934 \pm 0.006)$\\
(Gaussian) &$g_{\psi D^* D}=3.43 \pm 0.80$ GeV$^{-1}$&$-i(0.040 \pm 0.010)$&$-i(0.030 \pm 0.001)$\\
&$g_{\psi D^* D^*}=1.90 \pm 0.95$&&\\
&$M_{\psi}=4.01$ GeV&&\\
\hline
$F_{\Lambda}=(1+\frac{k^4}{\Lambda^4})^{-2}$&$g_{\psi DD}=21.6 \pm 4.4$&$(4.063 \pm 0.023)$&$(3.942 \pm 0.004)$\\
(Dipolar) &$g_{\psi D^* D}=3.05 \pm 0.49$ GeV$^{-1}$&$-i(0.057 \pm 0.010)$&$-i(0.052 \pm 0.010)$\\
&$g_{\psi D^* D^*}=2.00 \pm 0.91$&&\\
&$M_{\psi}=4.03$ GeV&&\\
\hline

\end{tabular}
\caption{The numerical values of the parameters of the model and the coordinates of the poles in the complex plane for two types of cutoff function and $\Lambda=0.42$ GeV.}
\end{table} 

 Naively, one is tempted to identify this enhancement and the corresponding pole with the puzzling state $Y(4008)$ and interprete it as a dynamically generated companion pole of $\psi(4040)$. However, a deeper analysis of the reaction $e^+ e^- \rightarrow \psi(4040) \rightarrow DD^* \rightarrow \pi^+ \pi^- J\psi$ can explain the formation of the state $Y(4008) $ as a $DD^*$ loop effect \cite{newpaper}, even without invoking the presence of a pole. 

\section{Conclusions}
We explored the well-established charm-anticharm vector resonance $\psi(4040)$ by using a quantum field approach. In particular, we studied its spectral function, which turns out to be not compatible with a standard Breit-Wigner shape due to a significant deformation in the energy region close to 3.9 GeV. Moreover, two poles emerge in the complex plane: an expected one for the resonance $\psi(4040)$  and a second one related to the left enhancement generated by $DD^*$ loops. Although a direct assignment of this additional pole to $Y(4008)$ is not possible, the study of the reaction  $e^+ e^- \rightarrow \psi(4040) \rightarrow DD^* \rightarrow \pi^+ \pi^- J\psi$ shows that $Y(4008)$ does not necessarly describe a genuine resonance, but is an effect of $DD^*$ quantum fluctuations.

  \begin{acknowledgement}
The authors thank P. Kovacs for cooperation and S. Coito for useful discussions and  acknowledge financial support from the Polish National Science
Centre (NCN) through the OPUS project no. 2015/17/B/ST2/01625.
\end{acknowledgement}
%
%
%

\end{document}